\documentclass[twocolumn,floatfix,aps, prx]{revtex4-1}

\usepackage{graphicx}
\usepackage{mathtools}
\usepackage{amsmath}
\usepackage{amssymb}
\usepackage{bm}
\usepackage{color}
\usepackage{comment}
\usepackage[FIGTOPCAP]{subfigure}
\usepackage[]{footmisc}
\usepackage{lineno}
\usepackage{dsfont}
\usepackage{soul,xcolor}
\usepackage[pdfborder={0 0 0}]{hyperref}

\newcommand{\ket}[1]{|#1\rangle}

\begin{document}

\title{Equilibrium current in a Weyl-semimetal - superconductor heterostructure}

\author{K.\ A.\ Madsen}
\author{P.\ W.\ Brouwer}
\author{M.\ Breitkreiz}
\email{breitkr@physik.fu-berlin.de}
\affiliation{Dahlem Center for Complex Quantum Systems and Fachbereich Physik, Freie Universit\" at Berlin, 14195 Berlin, Germany}

\date{\today}

\begin{abstract}

A heterostructure consisting of a magnetic Weyl semimetal and a conventional superconductor exhibits an equilibrium current parallel to the superconductor interface and perpendicular to the magnetization. Analyzing a minimal model, which as a function of parameters may be in a trivial magnetic insulator phase, a Weyl semimetal phase, or a three-dimensional weak Chern insulator phase, we find that the equilibrium current is sensitive to the presence of surface states, such as the topological Fermi-arc states of the Weyl semimetal or the chiral surface states of the weak Chern insulator. While there is a nonzero equilibrium current in all three phases, the appearance of the surface states in the topological regime leads to a reversal of the direction of the current, compared to the current direction for the trivial magnetic insulator phase. We discuss the interpretation of the surface-state contribution to the equilibrium current as a real-space realization of the superconductivity-enabled equilibrium chiral magnetic effect of a single chirality, predicted  to occur in bulk Weyl superconductors.

\end{abstract}
\maketitle

\section{Introduction}

A Weyl semimetal is a three-dimensional crystal with topologically protected nodal points in the band structure  \cite{Armitage2017, Yan2017, Burkov2017}. 
The nodes have a well-defined chirality and they appear in pairs, such that in total the sum of the chiralities vanishes \cite{Nielsen1983}. 
One  manifestation of chiral Weyl nodes and the associated chiral anomaly in crystals
is the existence of topologically protected surface states, which connect the projections of two Weyl nodes of opposite chirality on the surface band structure, in the form of two ``Fermi arcs'' located at opposite surfaces of the Weyl semimetal and moving in opposite directions.
 Another manifestation  is the chiral magnetic effect --- an external-magnetic-field induced current of Weyl Fermions directed parallel or antiparallel to the magnetic field depending on the  chirality --- which leads to unusual non-equilibrium transport properties of the crystal \cite{Kharzeev2014, Burkov2015, Xiong2015, Huang2015a, Reis2016}.  
 In equilibrium the chiral anomaly usually remains hidden, since the chiral currents must compensate each other, in agreement with general band-theoretic considerations \cite{Vazifeh2013}.

As was shown by O'Brien, Beenakker, and Adagideli \cite{Obrien2017} (see also Ref.\ \cite{Pacholski2020}), there is, however, a way to circumvent the compensation of chiral anomalies in equilibrium with the help of superconductivity. This is most easily seen in a minimal model of a magnetic Weyl semimetal with two Weyl nodes of opposite chirality and a superconducting s-wave pair potential. If the pair momentum is tuned to the momentum of one of the two Weyl nodes via a flux or a supercurrent bias, superconductivity is induced there and the Weyl node is gapped out, while the node of opposite chirality is left mostly unaffected. In an applied magnetic field, this unaffected chirality gives rise to an equilibrium current, as the opposite chirality is no longer available to carry the compensating current. Unfortunately, making a Weyl semimetal superconducting \cite{Kang2015, Qi2016, Zhu2018, Cai2019} meets the difficulty of a vanishing density of states at the Weyl nodes, which suppresses the critical temperature. Another obstacle, specifically  in the case of a magnetic Weyl semimetal considered in this work, is the competition with magnetism.
 
An alternative route to achieve superconducting phases in Weyl semimetals is to make use of the proximity-induced superconductivity in heterostructures by combining an otherwise non-superconducting Weyl semimetal (N) and a conventional superconductor (S) \cite{Wang2016, Bachmann2017, Shvetsov2020, Shvetsov2020a}. One prominent type of such heterostructures is the Josephson junction (SNS-heterostructure), which has been extensively studied theoretically exploring the influence of various 
 types of superconducting pairing mechanisms  \cite{Madsen2017, Bovenzi2017, Sinha2020,Dutta2020,Alidoust2020,Dutta2019,
Kim2016,Uddin2019,Chen2016,Chen2017,Alidoust2018,
Khanna2016, Kulikov2020,Khanna2017,Zhang2018},
 and has also been realized experimentally \cite{Kononov2019,Shvetsov2020,Choi2019,
 Huang2019,Shvetsov2018-1,Shvetsov2018-2}.  
 Other examples of similar heterostructures are NS-type 
 \cite{Howlader2019,Zhang2018a,Liu2017,Hou2017,
 Wang2016,Chen2013,Fang2018,Faraei2019,
 Shvetsov2020a,Grabecki2020,Naidyuk2018,
 Kononov2018,Aggarwal2017}, and NSN-type \cite{Breunig2019,Liu2018,Li2018,Li2019,Sinha2019} 
heterostructures.

\begin{figure}[t]
\includegraphics[width=0.6\columnwidth]{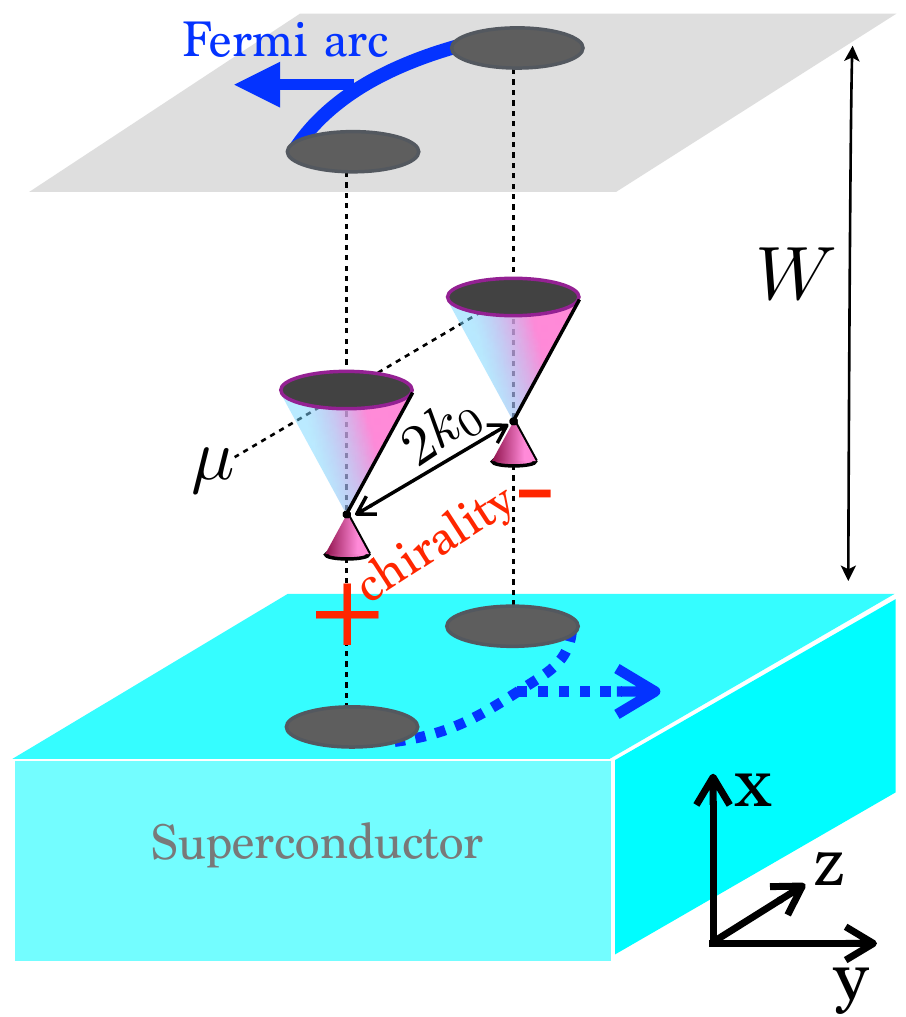}
\caption{Mixed momentum-/real-space illustration of the SN heterostructure considered in this article. It consists of a Weyl semimetal slab of a finite width $W$ with two Weyl nodes separated along the $k_z$ axis and counterpropagating Fermi arcs  on the top (solid blue) and 
bottom (dotted blue) surfaces. The Weyl semimetal slab borders on a superconductor (light blue) at the bottom surface and it is capped by a trivial insulator at the top surface. Because of this built-in spatial asymmetry of the heterostructure, the superconducting proximity effect acts asymmetrically on the two Fermi arcs.}
\label{fig:intro}
\end{figure}

While most of these studies investigate 
equilibrium currents that flow perpendicular to the superconductor - Weyl-semimetal interface, 
in this article we theoretically investigate the equilibrium current in a bilayer consisting of a Weyl semimetal and a single superconductor (SN bilayer), as illustrated in Fig.\ \ref{fig:intro}, for which the equilibrium current flows parallel to the interface. We consider a magnetic Weyl semimetal and a conventional s-wave superconductor, 
both are microscopically inversion-symmetric, 
so that inversion symmetry is broken only by the 
interface. To allow for a comparison between different phases, we consider a model for the normal region which, as a function of parameters, may be in a trivial magnetic insulator phase, Weyl semimetal phase, or a (three-dimensional) weak Chern insulator phase. We find a significant contribution to the equilibrium current from surface states (Fermi arcs in case of a Weyl semimetal, chiral surface states for the weak Chern insulator), which differs in sign and magnitude from the interfacial current of a trivial insulator \cite{Mironov2017a}. Although our minimal model shows a clear signature at the onset of the topological regime, the magnitude of the equilibrium current is non-universal, because for an inversion-symmetric Weyl semimetal the proximity superconductivity pairs electrons in the topological low-energy band with electrons in a non-topological high-energy band --- an effect known as ``chirality blockade'' \cite{Bovenzi2017}. For the minimal model we can isolate the singular contribution to the current from the Fermi-arc surface states by comparing equilibrium currents in a finite-width slab for a chemical potential inside and outside the finite-size gap of the Fermi-arc states at the Weyl node.

The contribution of topological surface states can be interpreted as the result of an effective charge renormalization of the  chiral surface modes at the SN interface \cite{Baireuther2017}, which leads to a 
disbalance with the counterpropagating surface modes 
of the opposite surface and results in a finite current. In this way, the idea of bulk superconductivity acting asymmetrically on chiral states in momentum space  \cite{Obrien2017, Pacholski2020} is transferred to proximitized superconductivity acting asymmetrically on chiral states in real space. In the former case the equilibrium current is carried by the disbalanced chiral Weyl Fermions in an external magnetic field, in the latter by the disbalanced chiral surface states at zero external magnetic field. 

This article is structured as follows: After introducing the minimal model for the SN heterostructure in Sec.~\ref{sec:model}, we calculate and discuss the equilibrium current in Sec.~\ref{sec:current}. We conclude in Sec.~\ref{sec:discussion}.

\section{Model \label{sec:model}}

We consider a bilayer consisting of a superconductor (S) and a normal region (N) of width $W$. We choose coordinates such that the $x$ axis is perpendicular to the superconductor interface and the superconductor interface is at $x=0$. The normal region corresponds to $0 < x < W$.

Depending on parameters in our model Hamiltonian, the normal region is a topologically trivial magnetic insulator, a magnetic Weyl semimetal, or a three-dimensional weak Chern insulator. At $x=W$ the normal region layer is capped by a non-magnetic trivial insulator. Below, we give lattice models for the Weyl semimetal, the superconductor, and the trivial insulator. To keep the notation simple, the lattice constant and $\hbar$ are set to unity.

\subsection{Normal region}

We model the normal region with the four-band Hamiltonian
\begin{align}
	H^{\rm (W)}(\bm{k}) =&\, t\tau_3 ( \sigma_1 \sin k_x + \sigma_2 \sin k_y)
	\nonumber \\ &\, \mbox{}
	+ m(\bm{k}) \tau_1 \sigma_0
	+ \beta \tau_0 \sigma_3 - \mu \tau_0 \sigma_0,
	\label{eq:H1}
\end{align}
with
\begin{align}
	m(\bm{k}) =&\, m_0 + t'(2 - \cos k_x - \cos k_y)
	\nonumber \\ &\, \mbox{} + t_z'(1-\cos k_z),
\end{align}
where the $\sigma_i$ and $\tau_i$, $i=0,1,2,3$ are Pauli matrices corresponding to spin and orbital degrees of freedom, respectively. (These include the identity matrices $\sigma_0$ and $\tau_0$.) Furthermore, $\mu$ is the chemical potential, $t$, $t'$, and $t_z'$ are hopping parameters, $m_0$ an orbital-selective on-site potential, and $\beta$ the exchange field, which is directed in the $z$ direction. For definiteness, all of these parameters are assumed to be positive. The Hamiltonian, shown in Eq.\ (\ref{eq:H1}), satisfies inversion symmetry,
\begin{equation}
  H^{\rm (W)}(\bm{k}) = \tau_1 H^{\rm (W)}(-\bm{k}) \tau_1, \label{invs}
\end{equation}
whereas time-reversal symmetry is broken by the exchange field. (Time-reversal symmetry is represented as $\sigma_2 K$, where $K$ is complex conjugation.) At zero chemical potential $\mu$, the Hamiltonian, see Eq.\ (\ref{eq:H1}), also satisfies a mirror antisymmetry,
\begin{equation}
  H^{\rm (W)}_{\mu=0}(k_x,k_y,k_z) = - \sigma_2 \tau_3 H^{\rm (W)}_{\mu = 0}(k_x,-k_y,k_z) \sigma_2 \tau_3. \label{eq:mirror}
\end{equation}
The Hamiltonian, given in Eq.\ (\ref{eq:H1}), resembles minimal models motivated by materials of the Bi$_2$Se$_3$ family \cite{Vazifeh2013}, where, however, for simplicity we omitted a term proportional to $\tau_3\sigma_3 \sin k_z$. 
[Such a term does not significantly alter the topological phases that we are going to study, but its absence makes the analysis more transparent. A term $\propto \tau_3 \sigma_3 \sin k_z$ preserves the inversion symmetry, Eq.\ (\ref{invs}), and the mirror antisymmetry, Eq.\ (\ref{eq:mirror}), at $\mu = 0$. We verified that our conclusions remain valid if we include this term.]

The eigenvalues of the Hamiltonian, Eq.\ (\ref{eq:H1}), can easily be calculated in closed form. For each momentum $\bm{k}$ there are four eigenvalues, labeled $\varepsilon_{\pm,\pm}$,
\begin{equation}
	\varepsilon_{\pm,\pm}(\bm{k}) = -\mu \pm \sqrt{t^2(\sin^2 k_x + \sin^2 k_y)
		+ (m(\bm{k}) \pm \beta)^2}. \label{eq:epsilon}
\end{equation}
The two bands with energy eigenvalues $\varepsilon_{\pm,+}(\bm{k})$ are completely gapped. The other two bands, which have energy eigenvalues $\varepsilon_{\pm,-}(\bm{k})$, may also be gapped or feature two Weyl nodes, depending on the value of the exchange field $\beta$. The Weyl-semimetal phase is found for
\begin{equation}
	m_0 < \beta < m_0 + 2 t_z'. \label{eq:weylregime}
\end{equation}
In this case, two Weyl nodes exist at $\bm{k} = (0,0,\pm k_0)$, with
\begin{equation}
  k_0 = 2\arcsin \sqrt{\frac{\beta - m_0}{2 t_z'}}.
  \label{eq:weylpoints}
\end{equation}

For $\beta \downarrow m_0$, one has $k_0 \to 0$: The two Weyl nodes merge at $k_z=0$ and gap out for $\beta < m_0$. Hence, for
\begin{equation}
0 < \beta < m_0 \label{eq:insulatorregime}
\end{equation}
the system becomes a trivial magnetic insulator. For $\beta \uparrow m_0 + 2 t_z'$, one has $k_0\to \pi$, and the Weyl nodes merge and gap out at the Brillouin zone boundary. For 
\begin{equation}
	\beta > m_0 + 2 t_z' \label{eq:weakregime}
\end{equation}
the system thus becomes a weak Chern insulator \cite{Hasan2010, Qi2011}, which has open surface-state contours extending over the whole Brillouin zone.

To prepare for the description of superconductor heterostructures using the Bogoliubov-de Gennes (BdG) formalism, we double the degrees of freedom by introducing holes with Hamiltonian $-\sigma_2 H^{\rm (W)}(-\bm{k})^* \sigma_2$. The resulting Bogoliubov-de Gennes Hamiltonian
\begin{equation}
  {\cal H}^{\rm (W)} = \begin{pmatrix} H^{\rm (W)} & 0 \\ 0 & -\sigma_2 H^{\rm (W)}(-\bm{k})^* \sigma_2 \end{pmatrix}
\end{equation}
has particle-hole symmetry,
\begin{equation}
  {\cal H}(\bm{k}) = -\nu_2 \sigma_2 {\cal H}(-\bm{k})^* \nu_2 \sigma_2,
  \label{eq:ph}
\end{equation}
where Pauli matrices $\nu_j$, $j=0,1,2,3$, represent the particle-hole degree of freedom.

\subsection{Heterostructure}

The normal region at $0 < x < W$ is embedded between a superconductor for $x < 0$ and a trivial insulator for $x > W$. The lattice Hamiltonians for the superconductor (S) and trivial insulator (I) in the Bogoliubov-de Gennes formulation are 
\begin{align}
	{\cal H}^{\rm (S)}(\bm{k}) =
	t \nu_3 \tau_3 \sigma_1 \sin k_x + \Delta \nu_1 \tau_0 \sigma_0, \\
	{\cal H}^{\rm (I)}(\bm{k}) = t \nu_3 \tau_3 \sigma_1 \sin k_x
        + m^{\rm (I)} \nu_3 \tau_1 \sigma_0,
\end{align}
where $\Delta > 0$ is the superconducting order parameter and $m^{\rm (I)} \to \infty$ the mass gap in the insulating region. Both the superconductor and the insulator satisfy inversion symmetry,
\begin{align}
  {\cal H}^{\rm (S,I)}(\bm{k}) =&\,
  \tau_1 {\cal H}^{\rm (S,I)}(-\bm{k}) \tau_1,
\end{align}
characteristic of superconducting order with even inversion parity, and time-reversal symmetry,
\begin{align}
  {\cal H}^{\rm (S,I)}(\bm{k}) =
  \sigma_2 {\cal H}^{\rm (S,I)}(-\bm{k})^* \sigma_2.
  \label{eq:inversion}
\end{align}  

To describe the heterostructure with an $x$-dependent Hamiltonian, we replace $k_x$ by $-i \partial_x$ and linearize the Hamiltonians ${\cal H}^{\rm (W)}$, ${\cal H}^{\rm (S)}$, and ${\cal H}^{\rm (I)}$ in $k_x$. In this way, we obtain the Hamiltonian
\begin{align}
  {\cal H} = -i t \nu_3 \tau_3 \sigma_1 \partial_x + {\cal M}(x),
  \label{eq:HBdG}
\end{align}
where
\begin{subequations}
\begin{align}
  {\cal M}(x) =&\, {\cal M}^{\rm (S)} \nonumber \\ \equiv&\,
  \Delta \nu_1 \sigma_0
  \label{eq:Ms}
\end{align}
for $x < 0$,
\begin{align}
  {\cal M}(x) =&\, {\cal M}^{\rm (W)} \nonumber \\ \equiv&\,
	t \nu_3 \tau_3 \sigma_2 \sin k_y
	\nonumber \\ &\, \mbox{} 
   + m(k_y,k_z) \nu_3 \tau_1 \sigma_0	
   + \beta \nu_0 \sigma_3 - \mu  \nu_3 \sigma_0,
\end{align}
for $0 < x < W$, and
\begin{align}
  {\cal M}(x) =&\, {\cal M}^{\rm (I)} \nonumber \\ \equiv&\,
  m^{\rm (I)} \nu_3 \tau_1 \sigma_0,
\end{align} \label{hetH}
\end{subequations}
for $x>W$, respectively. Here 
\begin{equation}
	m(k_y,k_z) = m_0 + t'(1-\cos k_y) + t_z'(1-\cos k_z)
\end{equation}
is the linearized mass term in the normal region.

\subsection{Block diagonalization, chirality, Fermi arcs}

A unitary transformation can be used to bring the Hamiltonian to a block-diagonal form. Labeling the two blocks 
by the parameter $\tau = \pm 1$, the transformation reads
\begin{align}
  {\cal H}_\tau  &=\big[ U\, {\cal H}\, U^\dagger\big]_\tau,\;\;\;\;\;\;
U = e^{i (\pi/4) \nu_0\tau_2\sigma_3}. \label{eq:bdiag}
\end{align}
The transformation acts non trivially only on the mass term, which transforms as
\begin{equation}
\Big[U \nu_3\tau_1\sigma_0 U^\dagger\big]_\tau =\tau \nu_3\sigma_3,
\end{equation} 
while the transformation of the other terms simply replaces $\tau_3$ by $\tau$. After the unitary transformation from Eq.\ (\ref{eq:bdiag}) the diagonal blocks of the Hamiltonian, Eq.\ (\ref{eq:HBdG}), then read
\begin{equation}
  {\cal H}_{\tau} = -i t \tau \nu_3 \sigma_1 \partial_x + {\cal M}_\tau(x),
  \label{eq:HtauBdG}
\end{equation}
with ${\cal M}_{\tau}(x) = {\cal M}^{\rm (S)}$, given by Eq.\ (\ref{eq:Ms}), for $x < 0$, ${\cal M}(x) = {\cal M}_{\tau}^{\rm (W)}$,
\begin{align}
	{\cal M}_\tau^{\rm (W)} =&\,
	t \tau \nu_3 \sigma_2 \sin k_y
	\nonumber \\ &\, \mbox{}
	+ m(k_y,k_z) \tau \mu_3 \sigma_3 - \mu \nu_3 \sigma_0 
	+ \beta \nu_0 \sigma_3 \label{eq:H12band}
\end{align} 
for $0 < x < W$, and ${\cal M}(x) = {\cal M}^{\rm (I)}$,
\begin{equation}
  {\cal M}^{\rm (I)}_{\tau} = m^{\rm (I)} \tau \nu_3 \sigma_3
\end{equation}
for $x > W$.
In the transformed basis, inversion, time-reversal, particle-hole conjugation, and the mirror antisymmetry shown in Eq.\ (\ref{eq:mirror}) are represented as $\tau_3 \sigma_3$, $\tau_2 \sigma_1 K$, $\nu_2 \tau_2 \sigma_1 K$, and $\sigma_2 \tau_3$, respectively.

After the unitary transformation, the Weyl nodes are found in the blocks $\tau=-1$ for electrons and $\tau=+1$ for holes, respectively. Expanding ${\cal H}_{\tau}^{\rm (W)}$ around the Weyl nodes in the form $\sum_i v_i \sigma_i (k_i - K_i)$, where $K_i$ is the node position, we can identify the chirality $\chi=\mathrm{sign} (v_1 v_2 v_3)$. For our convention that all model parameters are positive, $\chi=\mp$ for the node at $k_z=\pm k_0$ for both electrons and holes, as indicated for electrons in Fig.\ \ref{fig:intro}.

To find Fermi-arc surface states at the interface with the trivial insulator at $x = W$, we consider electron and hole eigenstates of the insulator that decay for $x>W$, taken at $x=W$,
\begin{align}
	\psi_{{\rm e}/{\rm h}}(W) =& a_{{\rm e}/{\rm h}} \begin{pmatrix} 1 \\ i \end{pmatrix} \label{eq:psiW} ,
\end{align}
with normalization coefficients $a_{{\rm e}/{\rm h}}$ that have to be determined separately. For the normal region $x < W$ we use the Ansatz
\begin{equation}
  \psi_{{\rm e}/{\rm h}}(x)=a_{{\rm e}/{\rm h}} \begin{pmatrix} 1 \\ i \end{pmatrix} e^{\alpha (x-W)}. \label{eq:psiansatz}
\end{equation}
The decay coefficient $\alpha > 0$ and the energy $\varepsilon$ can be found by insertion of the Ansatz of Eq.\ (\ref{eq:psiansatz}) into the Bogoliubov-de Gennes equation
\begin{equation}
	\big[{\cal H}_{\tau}^{\rm (W)}-\varepsilon \big] \begin{pmatrix} \psi_e(x) \\ \psi_h(x) \end{pmatrix} =0.
\end{equation}
For $\tau = -1$ we find an electron-like solution with $\alpha=\beta-m(k_y, k_z)$ and energy
\begin{equation}
  \varepsilon_{\rm e}(k_y,k_z) = -t \sin k_y - \mu.
  \label{eq:arcdisp}
\end{equation}
For $\tau = +1$, the solution is hole-like and has energy
\begin{equation}
  \varepsilon_{\rm h}(k_y,k_z) = -t \sin k_y + \mu.
\end{equation}
Both solutions move in the $y$ direction with velocity $v_{\rm F} = d \varepsilon_{{\rm e}/{\rm h}}/dk_y = - t \cos k_y$, as illustrated (for electrons) in Fig.\ \ref{fig:intro}. For small $k_y$ the condition $\alpha > 0$ is satisfied for $|k_z| < k_0$, {\em i.e.}, for $k_z$ between the two Weyl points.

\section{Equilibrium current \label{sec:current}}

Superconductor--normal-metal heterostructures with a magnetic N region are known to exhibit an equilibrium current in the direction of $\bm{E}\times\bm{B}$, where here the role of the time-reversal breaking (magnetic) field $\bm{B}$ is played by the exchange field (described by the term proportional to $\beta$ in ${\cal H}^{\rm (W)}$ and here pointing in the $z$ direction) and the role of the inversion-symmetry breaking (electric) field $\bm{E}$ is played by a confinement-potential gradient of the interface (here in the $x$ direction) \cite{Mironov2017a}. In our geometry we thus expect to find an equilibrium current in the $y$ direction.

\subsection{Scattering formulation}
\label{sec:scat}

We calculate the equilibrium current density $I_y$ as the derivative of the ground state energy $E$ to the vector potential $A_y$. The vector potential $A_y$ enters the Bogoliubov-de Gennes Hamiltonian ${\cal H}$ of Eq.\ (\ref{eq:HBdG}) via the standard substitution $k_y \to k_y - \nu_3 e A_y$. Then the equilibrium current $I_y$ is
\begin{align}
  I_y =&\, \frac{1}{2} \sum_{\tau} \int_{-\infty}^0 d\varepsilon \, \varepsilon \frac{\partial}{\partial A_y} \frac{d N_{\tau}(\varepsilon)}{d \varepsilon} \nonumber \\ =&\,
  - \frac{1}{2}  \sum_{\tau}  \int_{-\infty}^0 d\varepsilon \frac{\partial N_{\tau}(\varepsilon)}{\partial A_y},
\end{align}
where $d N_{\tau}(\varepsilon)/d\varepsilon$ is the density of states of the Hamiltonian ${\cal H}_{\tau}$ of Eq.\ (\ref{eq:HtauBdG}) and $N_{\tau}(\varepsilon)$ is the cumulative density of states.

The density of states $d N_{\tau}(\varepsilon)/d\varepsilon$ is a sum of delta-function contributions for $|\varepsilon| < \Delta$ and continuous otherwise. In principle, $d N_{\tau}(\varepsilon)/d\varepsilon$ may depend on $A_y$ in both the discrete and continuous parts of the spectrum \cite{beenakker1995}. To capture both contributions, we adopt a procedure used by Beenakker and one of us for the calculation of the Josephson effect in a chaotic quantum dot \cite{Brouwer1997}. Following Ref.\ \cite{Brouwer1997}, we determine $N_{\tau}(\varepsilon)$ by matching solutions of the Bogoliubov-de Gennes equation ${\cal H}_{\tau} \psi = \varepsilon \psi$ in the superconducting region $x < 0$ and in the normal region $x > 0$. To this end, we insert an ``ideal lead'' between the superconducting region at $x < 0$ and the normal region at $x > 0$, described by the Hamiltonian of Eq.\ (\ref{eq:HtauBdG}) with ${\cal M}_{\tau} = 0$. At the end of the calculation, the length of the ideal lead is sent to zero. In the ideal lead, the Bogoliubov-de Gennes equation is solved by the scattering states
\begin{equation}
	\psi_{\tau;\nu,\pm}(x) = e^{\pm i \varepsilon x/t} \ket{\nu,\pm \nu\tau}, \label{eq:special}
\end{equation}
where $\ket{\nu,\sigma}$ with $\nu$, $\sigma = \pm 1$ is an eigenspinor of $\nu_3$ at eigenvalue $\nu$ and of $\sigma_1$ at eigenvalue $\sigma$. The eigenstates $\psi_{\tau;\nu,+}$ and $\psi_{\tau;\nu,-}$ represent solutions moving in the positive and negative $x$ directions, respectively. The solutions with $\nu=1$ are electron-like; the eigenstates with $\nu=-1$ are hole-like. 

In the ideal-lead segment around $x=0$, the full solution of the Bogoliubiov-de Gennes equation is a linear combination of the scattering states given in Eq.\ (\ref{eq:special}),
\begin{equation}
  \psi_{\tau}(x) = \sum_{\nu} \left[ a_{\tau,\nu} \psi_{\tau;\nu,+}(x) + b_{\tau,\nu} \psi_{\tau;\nu,-}(x) \right].
\end{equation}
Viewing the coefficients $a_{\tau,\nu}$ and $b_{\tau,\nu}$ as amplitudes of quasiparticles incident on and reflected from the normal region, respectively, we may relate them via the scattering matrix $S_{\tau}(\varepsilon)$ of the normal region,
\begin{equation}
  \begin{pmatrix} b_{\tau,+} \\ b_{\tau,-} \end{pmatrix}
  = S_{\tau}(\varepsilon) \begin{pmatrix} a_{\tau,+} \\ a_{\tau,-} \end{pmatrix}.
  \label{eq:SmatrixW}
\end{equation}
(The dependence of $S_{\tau}(\varepsilon)$ on $k_y$ and $k_z$ is kept implicit.) When seen from the superconductor, the coefficients $a_{\nu}$ represent the reflected amplitudes, whereas the coefficients $b_{\nu}$ represent the incident amplitude, so that one has the relation
\begin{equation}
  \begin{pmatrix} a_{\tau,+} \\ a_{\tau,-} \end{pmatrix}
  = S_{\tau}^{\rm (S)}(\varepsilon) \begin{pmatrix} b_{\tau,+} \\ b_{\tau,-} \end{pmatrix},
  \label{eq:rs}
\end{equation}
where $S_{\tau}^{\rm (S)}(\varepsilon)$ is the scattering matrix of the superconducting region. Upon combining Eqs.\ (\ref{eq:SmatrixW}) and (\ref{eq:rs}), one finds that nontrivial solutions of the Bogoliubov-de Gennes equation exist only if
\begin{equation}
  \det[1 - S_{\tau}(\varepsilon) S_{\tau}^{\rm (S)}(\varepsilon)] = 0.
\end{equation}
Since $S_{\tau}(\varepsilon)$ and $S_{\tau}^{\rm (S)}(\varepsilon)$ are analytic functions of $\varepsilon$ in the upper half of the complex plane, we may directly obtain the cumulative density of states $N_{\tau}(\varepsilon)$ as \cite{Brouwer1997}
\begin{align}
  N_{\tau}(\varepsilon) =&\, -\frac{1}{\pi}
  \int \frac{dk_y dk_z}{(2 \pi)^2} 
  \mbox{Im}
  \left\{ \vphantom{\frac{1}{1}} \ln
  \det[1 - S_{\tau}(\varepsilon^+) S_{\tau}^{\rm (S)}(\varepsilon^+)
  \right. \nonumber \\ &\, \left. \mbox{}
  - \frac{1}{2}
  \ln [\det (S_{\tau}(\varepsilon)) ] - \frac{1}{2} \ln [\det (S_{\tau}^{\rm (S)}(\varepsilon))] \right\},
  \label{eq:Ntau}
\end{align}
where $\varepsilon^+ = \varepsilon + i \eta$, $\eta$ being a positive infinitesimal.
  
The second and third terms between the brackets in Eq.\ (\ref{eq:Ntau}) do not contribute to the current after integration to $k_y$. The first term in Eq.\ (\ref{eq:Ntau}) is analytic in the upper half of the complex plane and vanishes for $\mbox{Im}\, \varepsilon \to \infty$. Shifting the integration along the negative real axis to the positive imaginary axis, we then find
\begin{align}
  I_y =&\, \int \frac{dk_z}{2 \pi} {\cal I}_y(k_z), \label{eq:Itotal}
\end{align}
where
\begin{align}
  {\cal I}_y(k_z) =&\,
  - \frac{1}{2 \pi} \sum_{\tau} \int \frac{dk_y}{2 \pi}
  \mbox{Re}\, \int_0^{\infty} d\omega \nonumber \\ &\, \mbox{} \times
  \frac{\partial}{\partial A_y}
  \ln
  \det[1 - S_{\tau}(i\omega) S_{\tau}^{\rm (S)}(i \omega)].
  \label{eq:Iscat}
\end{align}

Under particle-hole conjugation, the basis state $\psi_{\tau;\nu,\pm}(x)$ of Eq.\ (\ref{eq:special}) is mapped to $\mp \psi_{-\tau;-\nu,\pm}(x)$, while simultaneously inverting $\varepsilon \to -\varepsilon$ and $k_{y,z} \to - k_{y,z}$, and vice versa. For this choice of the scattering states, particle-hole symmetry imposes the condition
\begin{equation}
  S_{\tau}(\varepsilon;k_y,k_z) = -\nu_1 S^*_{-\tau}(-\varepsilon;-k_y,-k_z) \nu_1.
\end{equation}
Calculating the scattering matrix $S^{\rm (S)}$ of the superconductor one obtains
\begin{equation}
  S_{\tau}^{\rm (S)}(\varepsilon) = e^{-i \gamma(\varepsilon)} \nu_1,\ \
  \gamma = \arccos(\varepsilon/\Delta),
  \label{eq:SAndreev}
\end{equation}
which is the standard result for Andreev reflection off an $s$-wave spin-singlet superconductor \cite{andreev1964}. The scattering matrix $S_{\tau}(\varepsilon)$ of the normal region is diagonal with respect to the particle-hole index $\nu$,
\begin{align}
  S_{\tau}(\varepsilon;k_y,k_z) = \begin{pmatrix}
    r_{\tau}(\varepsilon;k_y,k_z) & 0 \\ 0 & -r_{-\tau}(-\varepsilon;-k_y,-k_z)^*
  \end{pmatrix},
  \label{eq:Sdiag}
\end{align}
where $r_{\tau}(\varepsilon;k_y,k_z)$ is the reflection amplitude for electron-like quasiparticles. Inserting Eqs.\ (\ref{eq:SAndreev}) and (\ref{eq:Sdiag}) into Eq.\ (\ref{eq:Iscat}) and performing a partial integration to $k_y$, we find
\begin{align}
  \label{eq:Iscat2}
  {\cal I}_y(k_z) =&\, \frac{2 e}{\pi} \int \frac{dk_y}{2 \pi}
  \mbox{Re}\, \int_0^{\infty} d\omega 
  \frac{\partial r_+(i \omega;k_y,k_z)}{\partial k_y}
  \\ \nonumber &\, \mbox{} \times
  \frac{r_-(i \omega;-k_y,-k_z)^*}{e^{2 i \gamma(i \omega)} + r_+(i \omega;k_y,k_z) r_-(i \omega;-k_y,-k_z)^*}.
\end{align}

Because of the mirror antisymmetry  at $\mu = 0$ given in Eq.\ (\ref{eq:mirror}), the reflection amplitudes satisfy $r_{\tau}(\varepsilon;k_y,k_z) = r_{\tau}(\varepsilon;-k_y,k_z)^*$,
 from which it follows that the current vanishes 
 at $\mu = 0$. We use this feature of our model to focus 
 our calculation on the derivative 
 $d {\cal I}_y(k_z)/d\mu$ at small $\mu$. 

\subsection{Reflection amplitudes of normal region}

We calculate the reflection amplitude $r_{\tau}$ by expressing it in terms of the reflection and transmission amplitudes $r_{\tau}^{\rm (W)}$, $r_{\tau}'{}^{\rm (W)}$, $t_{\tau}^{\rm (W)}$, and $t_{\tau}'{}^{\rm (W)}$ of the normal region $0 < x < W$ and the reflection phase $i \tau$ of the insulator at $x > W$,
\begin{equation}
  r_{\tau} = r^{\rm (W)}_{\tau} + \frac{i \tau t_{\tau}'{}^{\rm (W)} t_{\tau}^{\rm (W)}}{1 - i \tau r_{\tau}'{}^{\rm (W)}}.
\end{equation}
In this notation, the unprimed amplitudes $r_{\tau}^{\rm (W)}$ and $t_{\tau}^{\rm (W)}$ refer to reflection and transmission from the normal region for particles incident at the interface with the superconductor (S), whereas the primed amplitudes $r_{\tau}'{}^{\rm (W)}$ and $t_{\tau}'{}^{\rm (W)}$ are for particles incident at the interface with the trivial insulator (I).
Solving the scattering problem with the Hamiltonian of Eq.\ (\ref{eq:HtauBdG}), we find the explicit expressions
\begin{align}
  r_{\tau}^{\rm (W)}(\varepsilon;k_y,k_z) =&\, r_{\tau}'{}^{\rm (W)}(\varepsilon;-k_y,-k_z) \nonumber \\ =&\, i \tau \frac{m(k_y,k_z) + \beta \tau - i t \tau \sin k_y}{t \kappa_{\tau} \coth(\kappa_{\tau} W) - i (\varepsilon + \mu)}, \label{eq:rtauW}\\
  t_{\tau}^{\rm (W)}(\varepsilon;k_y,k_z)  =&\, t_{\tau}'{}^{\rm (W)}(\varepsilon;-k_y,-k_z) \nonumber \\ =&\, \frac{t \kappa_{\tau}/\sinh(\kappa_{\tau} W)}{t \kappa_{\tau} \coth(\kappa_{\tau} W) - i(\varepsilon + \mu)},
\end{align}
where we abbreviated
\begin{align}
  \kappa_{\tau}^2 t^2 = d_{\tau}(k_x,k_y)^2 - (\varepsilon + \mu)^2,
\end{align}
with
\begin{equation}
  d_{\tau}(k_y,k_z) = \sqrt{t^2 \sin^2 k_y + (\beta + \tau m(k_y,k_z))^2}
  \label{eq:dtau}
\end{equation}
the gap in the $k_z$-dependent spectrum of the Hamiltonian shown in Eq.\ (\ref{eq:HtauBdG}), see Eq.\ (\ref{eq:epsilon}). The symmetry relation between primed and unprimed reflection and transmission amplitudes is a consequence of the inversion symmetry from Eq.\ (\ref{eq:inversion}).

To evaluate the $k_z$-resolved current density ${\cal I}_y(k_z)$, it is convenient to consider the three-dimensional Hamiltonian ${\cal H}(k_x,k_y,k_z)$ as family of two-dimensional Hamiltonians ${\cal H}(k_x,k_y)$ that parametrically depend on $k_z$. The two-dimensional Hamiltonian ${\cal H}^{\rm (W)}(k_x,k_y)$ describes a trivial (two-dimensional) insulator if $\beta < m_0$ or if $m_0 < \beta < m_0 + 2 t_z'$ and $|k_z| > k_0$, see Eqs.\ (\ref{eq:weylregime})--(\ref{eq:insulatorregime}). It describes a (two-dimensional) topologically nontrivial Chern insulator if $m_0 < \beta < m_0 + 2 t_z'$ and $|k_z| < k_0$ or if $\beta > m_0 + 2 t_z'$.

For the calculation of the equilibrium current $I_y$, we find it convenient to parameterize the reflection amplitudes $r_{\tau}^{\rm (W)}$, and $r_{\tau}'{}^{\rm (W)}$ in terms of the transmission coefficient $T_{\tau} = |t_{\tau}^{\rm (W)}|^2$ and the phase shifts $\phi_{\tau}$ and $\phi_{\tau}'$,
\begin{align}
  r_{\tau}^{\rm (W)} =&\, i \tau \sqrt{1 - T_{\tau}} e^{i \phi_{\tau}},\ \
  \nonumber \\
  r_{\tau}'{}^{\rm (W)} =&\, i \tau \sqrt{1 - T_{\tau}} e^{i \phi_{\tau}'}.
  \label{eq:rparam}
\end{align}
Expressions for the reflection phases $\phi_{\tau}$ and $\phi_{\tau}'$ can be obtained from Eq.\ (\ref{eq:rtauW}). For small $k_y$, $\varepsilon$, and $\mu$, the reflection phase $\phi_+$ of the high-energy band is well approximated by
\begin{align}
  \phi_{+}(k_y,k_z) =&\, \phi_{+}'(-k_y,-k_z)
  \nonumber \\ \approx&\, (\varepsilon + \mu - k_y t)/d_+.
  \label{eq:rapproxplus}
\end{align}
The approximations for the reflection phase for the low-energy band for small $k_y$, $\varepsilon$, and $\mu$ are different for the trivial regime $\beta < m_0$ or $|k_z| > k_0$ and the topological regime $\beta > m_0 + 2 t'_z$ or $|k_z| < k_0$,
\begin{align}
  \label{eq:rapproxmin}
  \phi_{-}(k_y,k_z) =&\, \phi_{-}'(-k_y,-k_z) \\  \nonumber \approx &\,
  \left\{ \begin{array}{ll}
    (\varepsilon + \mu + k_y t)/d_- & \mbox{trivial}, \\
    \pi + (\varepsilon + \mu - k_y t)/d_-& \mbox{topological}.
  \end{array} \right.
\end{align}
The fact that $\phi_- = \pi$ at $k_y=0$ in the topological case is what causes the appearance of the Fermi-arc surface states near $k_y = 0$. With the parameterization defined in Eqs.\ (\ref{eq:rparam}), the reflection amplitude $r_{\tau}(\varepsilon;k_y,k_z)$ reads
\begin{equation}
  r_{\tau} = i \tau e^{i \phi_{\tau}}
  \frac{e^{i \phi_{\tau}'} + \sqrt{1 - T_{\tau}}}
       {e^{i \phi_{\tau}'} \sqrt{1 - T_{\tau}} + 1}.
       \label{eq:rtau}
\end{equation}

\subsection{$k_z$-resolved current density for large $W$}
\label{sec:kz}

 We will now discuss the $k_z$-resolved current ${\cal I}_y(k_z)$ well inside the trivial and topological regimes, so that the two-dimensional Hamiltonian ${\cal H}^{\rm (W)}(k_x,k_y)$ describes a gapped phase  with a gap magnitude
  on the order of the band width. The case that $k_z$ is in the vicinity of $k_0$ will be addressed in Subsec.\ \ref{sec:weyl}.

For our calculation of ${\cal I}_y(k_z)$ we assume that the width $W$ of the normal region is much larger than the lattice spacing (which is set to one). The energy scale corresponding to the inverse width, $t/W$, the pair potential $\Delta$, and the chemical potential $\mu$ are considered to be much smaller that the band width $t \sim t' \sim t_z' $. The energy difference of the high- and low-energy bands, $2 m_0$, is considered to be on the order of the band width.  

With this hierarchy of energy and length scales, the energy dependence of the reflection amplitudes of the normal region may typically be neglected when compared to the energy dependence of the phase shift $\gamma$ for Andreev reflection from the superconductor. Also, one has $\kappa_{\tau} W \gg 1$, so that transmission is exponentially suppressed, $T_{\tau} \downarrow 0$. Assuming continuity of the current with $T_{\tau} \downarrow 0$, which we discuss in more 
detail in App.\ \ref{app:b}, we may set
\begin{equation}
  r_{\tau}(i \omega;k_y,k_z) = i \tau e^{i \phi_{\tau}(k_y,k_z)}, \label{eq:romegaapprox}
\end{equation}
where the reflection phase $\phi_{\tau}(k_y,k_z)$ of the normal region is evaluated at $\varepsilon = 0$. This approximation breaks down if $e^{i \phi_{\tau}'} = - 1$, because then the denominator in Eq.\ (\ref{eq:rtau}) vanishes for $T_{\tau} \downarrow 0$, which occurs if a Fermi-arc state at the surface at $x = W$ crosses the Fermi level. This case will be discussed in Subsec.\ \ref{A2}.
With the approximation from Eq.\ (\ref{eq:romegaapprox}), the $\omega$-integration in Eq.\ (\ref{eq:Iscat2}) may then be performed, with the result
\begin{align}
  {\cal I}_y(k_z) =&\, - \frac{e \Delta}{2} \int \frac{dk_y}{2 \pi}
  \frac{\partial \phi_+}{\partial k_y}
  s(\phi)
  \sin (\phi/2),
  \label{eq:Iphi}
\end{align}
where
\begin{equation}
  \phi(k_y,k_z) = \phi_+(k_y,k_z) - \phi_-(-k_y,-k_z)
\end{equation}
and $s(\phi) = \mbox{sign}\, \cos(\phi/2)$.

Effectively, the approximations used to derive Eq.\ (\ref{eq:Iphi}) from the general result of Eq.\ (\ref{eq:Iscat2}) amount to restricting to contributions from the discrete part of the Andreev spectrum. (This approximation is known as the ``short-junction limit'' in the context of the Josephson effect.) To show that Eq.\ (\ref{eq:Iphi}) represents the contribution from the discrete part of the Andreev spectrum, we note that, if we neglect the energy dependence of the reflection amplitudes from the normal region, Andreev bound states appear at discrete energies $\varepsilon_{\pm}(k_y,k_z)$ satisfying the quantization condition
\begin{equation}
  e^{-i 2 \gamma(\varepsilon_{\pm})} e^{i \phi_+(k_y,k_z)} e^{i \phi_-(-k_y,-k_z)} = 1.
\end{equation}
Solving for $\varepsilon_{\pm}(k_y,k_z)$, one finds
\begin{equation}
  \varepsilon_{\pm}(k_y,k_z) = \pm \Delta \cos(\phi/2). \label{eq:varepsilon}
\end{equation}
The current associated with a single Andreev level is $\partial \varepsilon_{\pm}(k_y,k_z)/\partial A_y$. To find the total current we integrate over the contributions from all Andreev levels with energy $\varepsilon_{\pm}(k_y,k_z) < 0$,
\begin{align}
 {\cal I}_y(k_z) =&\, \frac{1}{2} \sum_{\pm}
  \int \frac{dk_y}{2 \pi} \frac{\partial \varepsilon_{\pm}}{\partial A_y}
  \Theta(-\varepsilon_{\pm}),
\end{align}
where the Heaviside function $\Theta(x) = 1$ if $x > 0$ and $0$ otherwise. Upon substitution of Eq.\ (\ref{eq:varepsilon}) for $\varepsilon_{\pm}$, one recovers Eq.\ (\ref{eq:Iphi}).

To find the derivative $d {\cal I}_y(k_z)/d \mu$
 (recall that ${\cal I}_y(k_z) = 0$ for $\mu = 0$, 
see the discussion at the end of Subsec.\ \ref{sec:scat}) we observe that from Eq.\ (\ref{eq:rtauW}) we have
\begin{equation}
  \frac{\partial \phi_{\tau}}{\partial \mu} = \frac{1}{d_{\tau}}, \label{eq:phipm}
\end{equation}
where the gap $d_{\tau}(k_y,k_z)$ was defined in Eq.\ (\ref{eq:dtau}). For the $\mu$-derivative of the $k_z$-resolved current ${\cal I}_y(k_z)$ we then find a ``regular'' contribution and a ``singular'' contribution, which follows from the derivative of the discontinuity of the step function $s(\phi)$ at $\phi = \pi$ ($\mbox{mod}\, 2 \pi$),
\begin{equation}
  \frac{d {\cal I}_y(k_z)}{d \mu} =
  \frac{d {\cal I}_y(k_z)}{d \mu}^{\rm (r)} +
  \frac{d {\cal I}_y(k_z)}{d \mu}^{\rm (s)},
  \label{eq:jy}
\end{equation}
with
\begin{align}
  \frac{d{\cal I}_y(k_z)}{d \mu}^{\rm (r)} =&\,
  - \frac{e \Delta}{4} \int \frac{dk_y}{2 \pi} 
  \left[
  \left( \frac{1}{d_+} - \frac{1}{d_-} \right)
  \frac{\partial \phi_+}{\partial k_y}
  \cos \frac{\phi}{2}
  \right. \nonumber \\ &\, \left. \ \ \ \ \mbox{}
  -
  \frac{2}{d_+^2} \frac{\partial d_+}{\partial k_y}
  \sin \frac{\phi}{2} \right]
  s(\phi), \label{eq:jyr}\\
  \frac{d {\cal I}_y(k_z)}{d \mu}^{\rm (s)} =&\, e \Delta
  \int \frac{dk_y}{2 \pi}
  \frac{\partial \phi_+}{\partial k_y}  
  \left( \frac{1}{d_+} - \frac{1}{d_-} \right)
  \delta(\phi - \pi), \label{eq:jys}
\end{align}
where the delta function should be periodically extended with period $2 \pi$.
In the limit of a large exchange field $\beta$, $d_-$
is much smaller than $d_+$ and one may further
approximate $d{\cal I}_y(k_z)/d\mu$ by restricting 
to the terms inversely proportional to $d_-$.

\begin{figure}
\includegraphics[width=\columnwidth]{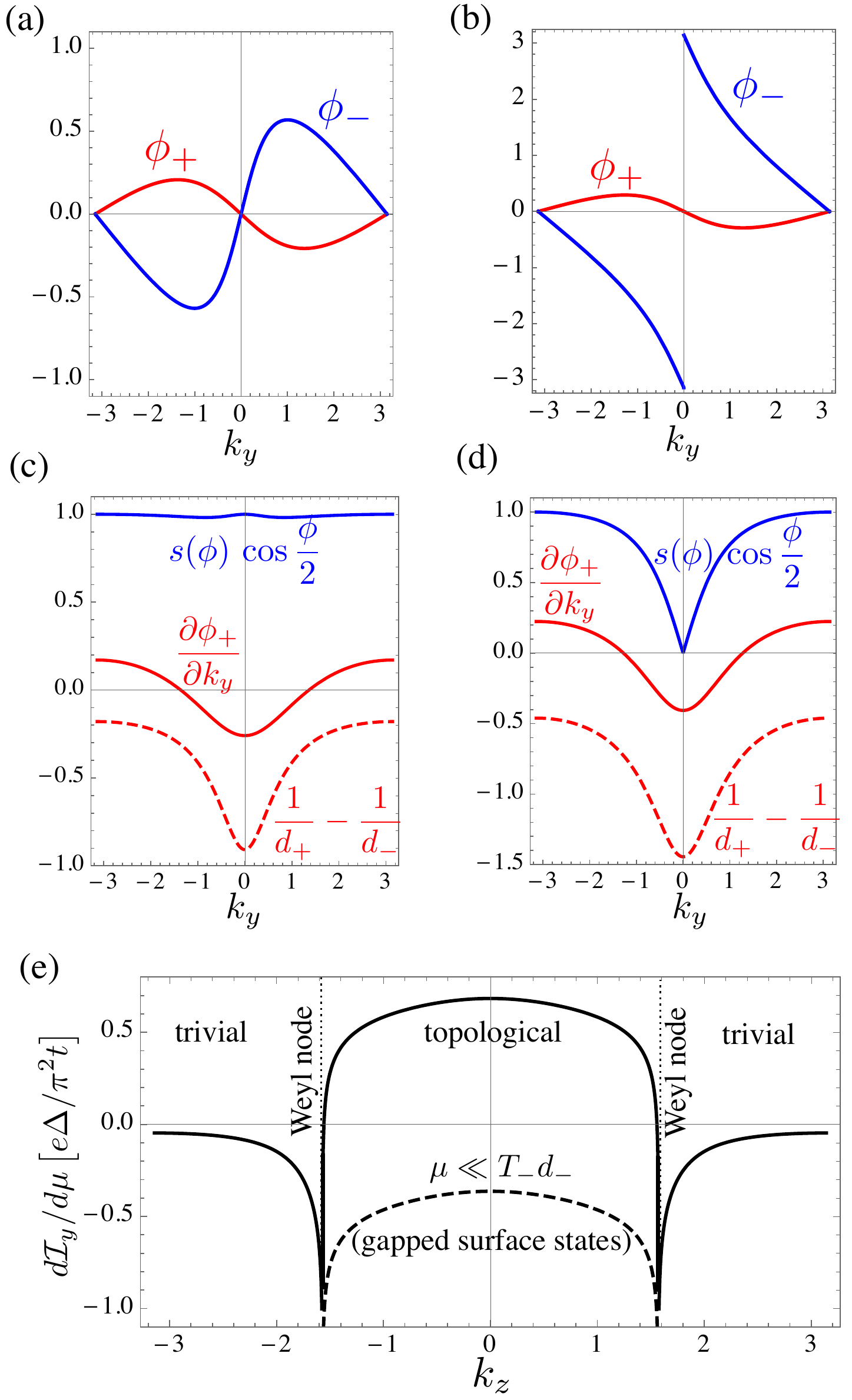}
\caption{(a) and (b): Reflection phases $\phi_{\pm}( k_y,k_z)$ at chemical potential $\mu \to 0$ and energy $\varepsilon = 0$ (after first taking the limit $W \to \infty$) for parameter choices corresponding to the trivial (a) and topological (b) regimes. (c) and (d): Factors $s(\phi) \sin(\phi/2)$ (blue), $(t/d_+ - t/d_-)$ (red, dashed) and $\partial \phi_+/\partial k_y$ (red, solid) for the same parameter choices as in (a) and (b), respectively. (e): $k_z$-resolved equilibrium current ${\cal I_y}(k_z)$ as a function of $k_z$ from Eq.\ (\ref{eq:jy}) (solid curve). The sign of the current changes if $k_z$ goes from the topological region ($k_z$ between the Weyl nodes at $\pm k_0$) to the trivial region. The dashed line shows the result at ultrasmall chemical potential within the finite-size gap of surface states, see Eq.\ (\ref{eq:jyfinite}). The parameters are $m_0=0.5\, t$, $\beta=1.5\, t$, $t= t'= t_z'= 1$. In panels (a) and (c) we further set $k_z=1$; in panels (b) and (d) we set $k_z=2.6$. }
\label{fig:analytics}
\end{figure}

On the basis of Eqs.\ (\ref{eq:jyr}) and (\ref{eq:jys}) we can compare $d{\cal I}_y(k_z)/d\mu$ in the trivial and topological regimes. The phases $\phi_+$ and $\phi_-$ are shown vs.\ $k_y$ for typical model parameters in  Fig.\ \ref{fig:analytics}(a) and (b). In the topologically trivial case, generically both $\phi_+$ and $\phi_-$ have a weak $k_y$-dependence and $\phi$ remains close to zero. In this case, the singular contribution $[d{\cal I}_y(k_z)/d\mu]^{\rm (s)}$ is absent. Considering the ``regular'' contribution (\ref{eq:jyr}), we see that the dominant contribution to the total equilibrium current $ I_y $ comes from regions in which the gap $d_{-}$ is smallest, which is in the vicinity of the Weyl points, {\em i.e.}, for $|k_z| \downarrow k_0$. The sign of the equilibrium current is determined by the derivative $d\phi_+/dk_y$ near $k_y = 0$.

In the topological case, as a result of the band inversion from the sign change of $\beta - m(k_y,k_z)$, the phase $\phi_-$ decreases by $2 \pi$ upon going from $k_y = -\pi$ to $k_y = \pi$. Hence, the singularity in the integrand at $\phi = \pi$ ($\mbox{mod}\, 2 \pi$) cannot be avoided. This gives rise to the singular contribution $[d{\cal I}_y(k_z)/d\mu]^{\rm (s)}$ of Eq.\ (\ref{eq:jys}). Since $\phi_-$ is close to $\pi$ in the vicinity of $k_y=0$, the integrand in Eq.\ (\ref{eq:jys}) has support precisely where the derivative $\partial \phi_+/\partial k_y$ is maximal, see Fig.\ \ref{fig:analytics}(c).  As a consequence, in the topological regime, the total current $d{\cal I}_y (k_z)/d\mu$ has larger magnitude and opposite sign when compared to the trivial regime, see Fig.\ \ref{fig:analytics}(e). 

To obtain an explicit expression for a special parameter choice well inside the topological regime, one may consider $k_z=0$ and $\beta=m_0+t'$, $t'=t$, in which case $\kappa_-= 1$ and $\phi_-(k_y,0) \approx \pi - k_y$ for all $k_y$. Additionally assuming a large gap $d_+\approx \beta+m_0\gg t$, so that $\phi_+(k_y,0) \approx -(t/(m_0+\beta)) \sin k_y$, the current becomes 
\begin{equation}
  \frac{d{\cal I}_y(0)}{d\mu} \approx \frac{2 e\Delta}
 {3\pi  (\beta+m_0)}.\label{tcur}
\end{equation}
For the trivial case we consider the
leading-order term in $\beta/t$, 
since the current vanishes at $\beta=0$, and take $m_0=t=t'$ and $k_z=0$, which gives  
\begin{equation}
  \frac{d{\cal I}_y(0)}{d\mu} \approx - \frac{e\Delta\beta}{12\pi t^2}.
    \label{ttcur}
\end{equation}
Comparing Eqs.\ (\ref{tcur}) and (\ref{ttcur}) also shows the opposite signs of the equilibrium current in the two regimes.

\subsection{Finite-size effects \label{A2}}

For small transmission coefficient $T_-$ of the low energy band, the presence of the Fermi-arc states at the interface with the trivial insulator at $x=W$ causes a narrow resonance in the reflection amplitude $r_-(\varepsilon;k_y,k_z)$. This resonance occurs, when the denominator in Eq.\ (\ref{eq:rtau}) is approximately zero, $e^{i \phi_{\tau}'} \approx -1$. In this case, the assumption that the energy dependence of $r_-(\varepsilon;k_y,k_z)$ can be neglected when compared to the energy dependence of the Andreev reflection phase $e^{-i \gamma(\varepsilon)}$ is obviously violated, despite the fact that the gap $d_- \gg \Delta$.

For the minimal model we consider in this article, this issue affects the topological regime $\beta > m_0$, $|k_z| < k_0$ only. Here we consider the case of small $\mu \ll t$, so that the resonance appears in the vicinity of $k_y = 0$. For small transmission coefficient $T_{-}$, the full reflection amplitude $r_-$ of Eq.\ (\ref{eq:rtau}) may then be well approximated as
\begin{equation}
  r_- = -i e^{i \phi_-} w(k_y t + \varepsilon + \mu),
\end{equation}
with
\begin{equation}
  w(\varepsilon) =
  \frac{2 \varepsilon - i T_- d_-}{2 \varepsilon + i T_- d_-}. \label{eq:rminfinitesize}
\end{equation}
Since $w(k_y t + i \omega + \mu) \approx 1$ if $|k_y t + i \omega + \mu| \gtrsim T_- d_-$, the presence of the factor $w(k_y t + i \omega + \mu)$ has little effect on the integrand in Eq.\ (\ref{eq:Iscat2}) in the limit of small transmission $T_-$ if $\mu \gg T_- d_-$, except for a small integration region around $k_y t \approx - \mu$ and $\omega \lesssim T_- d_-$. Because of the smallness of the integration region in which $w$ significantly differs from unity, the net finite-size effect on $d {\cal I}_y(k_z)/d\mu$ after integration over $k_y$ and $\omega$ is small and goes to zero if $T_- \to 0$. For $\mu \lesssim T_- d_-$ this conclusion cannot be drawn, however, because the singularity in the fraction in Eq.\ (\ref{eq:rminfinitesize}) coincides  with the 
 singularity
of  the integrand in  $d {\cal I}_y(k_z)/d\mu$, which led to
the singular contribution shown in Eq.\ \eqref{eq:jys}.

To analyze this limit of ``ultrasmall'' chemical potential $\mu$ in further detail, we observe that the singular contributions of the integration in Eq.\ (\ref{eq:Iscat2}) from the vanishing of the denominator and from the finite-size factor $w(k_y t + i \omega + \mu)$ are limited to a small interval $-\delta < k_y < \delta$ around $k_y = 0$, where $\delta \ll 1$ may be chosen large enough that $w(\pm \delta t + \mu + i \omega) \approx 1$. It follows that the ``regular'' contribution of Eq.\ (\ref{eq:jyr}) to $d{\cal I}_y(k_z)/d\mu$, which is associated with momenta $k_y$ outside this interval, is unaffected by the finite-size effects. On the other hand, as we show in detail in App. \ref{app:a}, upon inclusion of the finite-size effects the
 integrand of the singular contribution $d{\cal I}_y(k_z)^{\rm (s)}/d\mu$ is multiplied by a negative factor $-(d_++d_-)/(d_+-d_-)$, when compared to the result given in Eq.\ (\ref{eq:jys}) for $\mu \gg T_- d_-$. Hence for ultrasmall chemical potential $\mu \ll T_- d_-$ we find
\begin{equation}
  \frac{d {\cal I}_y(k_z)}{d \mu} =
  \frac{d {\cal I}_y(k_z)}{d \mu}^{\rm (r)} +
  \frac{d  {\cal I}_y(k_z)}{d \mu}^{{\rm (s)}},
  \label{eq:jyfinite}
\end{equation}
with $[d{\cal I}_y(k_z)/d\mu]^{\rm (r)}$ given by Eq.\ (\ref{eq:jyr}) and
\begin{align}
  \frac{d {\cal I}_y(k_z)}{d \mu}^{\rm (s)} =&\,
  e \Delta
  \int \frac{dk_y}{2 \pi}
  \frac{\partial \phi_+}{\partial k_y}  
  \left( \frac{1}{d_+} + \frac{1}{d_-} \right)
  \delta(\phi - \pi). \label{eq:jysfinite}
\end{align}
The sign change of the singular contribution leads to a significant {\em reduction} of the equilibrium current in the case of an ultrasmall chemical potential $\mu \ll T_- d_-$, when compared to the case $\mu \gg T_- d_-$.

To obtain an order-of-magnitude estimate, we again set $k_z = 0$ and consider the well-established topological regime $\beta=m_0+t'$, $t'=t$, $k_z=0$, $\beta+m_0\gg 1$, for which we find, that
\begin{align}
  \frac{d {\cal I}_y(0)}{d \mu} \approx -
 \frac{e \Delta}{3\pi  (\beta+m_0)} \label{diff}
\end{align}
if $\mu \ll T_- d_-$. Comparison to Eq.\ (\ref{tcur}) shows that at ultrasmall chemical potential the equilibrium current is approximately $-1/2$ times the current at finite $\mu$.

Physically, the energy $\sim T_- d_- \sim t \, e^{-2W}$ that separates the regimes of ``ultrasmall'' and ``finite'' $\mu$, is associated with the finite-size gap of the Fermi-arc surface states,  whose wavefunctions decay exponentially
away from the surfaces. Based on our result that in the topological regime the equilibrium current 
is strongly modified when the chemical potential is inside this finite-size gap, we interpret the difference between the finite-$\mu$ and ultrasmall-$\mu$ limits as the contribution of the topological surface states to $d{\cal I}_y/d\mu$. The difference between the large-$\mu$ and small-$\mu$ limits involves the singular contribution $[d {\cal I}_y/d\mu]^{\rm (s)}$ only. In the 
well-established topological regime  the surface-state
contribution assumes the value $2 [d {\cal I}_y/d\mu]^{\rm (s)}$, with $ [d {\cal I}_y/d\mu]^{\rm (s)}$ given in Eq.\ \eqref{eq:jys}.

\subsection{Total current density}
\label{sec:weyl}

The full equilibrium current density $I_y$ involves the integral of ${\cal I}_y(k_z)$ over $k_z$. The $k_z$-resolved current density ${\cal I}_y(k_z)$ is calculated in Sec.\ \ref{sec:kz}, for the case that the normal region is gapped at momentum $k_z$ and that the gap $d_{\tau} \gg \Delta$. This condition is no longer satisfied for the low-energy band if $k_z$ is in the immediate vicinity of the Weyl points, because $d_- \to 0$ there.

That the results of Sec.\ \ref{sec:kz} cease to be valid if $d_-$ becomes small in comparison to $\Delta$ is also reflected in the expression in Eq.\ (\ref{eq:jy}) for $d{\cal I}_y(k_z)/d\mu$, which diverges $\propto \Delta/d_-$ if $d_-/\Delta \to 0$. This divergence should be cut off for $d_- \sim \Delta$. To see this, we evaluate $d{\cal I}_y(k_z)/d\mu$ in the opposite limit $d_- \ll \Delta$, in which we may neglect the energy dependence of the Andreev reflection phase $e^{-i \gamma(\varepsilon)}$ and of the reflection amplitude $r_+$ of the high-energy band, but keep the full energy dependence of the reflection amplitude $r_-$ of the low-energy band.

Starting point of our calculation is Eq.\ (\ref{eq:Iscat2}). Since $r_-$ depends on energy $\varepsilon$ and chemical potential $\mu$ through the combination $\varepsilon + \mu$ only, upon analytic continuation $\varepsilon \to i \omega$, one has $\partial r_-^*/\partial \mu = i \partial r_-^*/\partial \omega$. When calculating $d {\cal I}_y(k_z)/d\mu$, the integrand in Eq.\ (\ref{eq:Iscat2}) then is a total derivative to $\omega$ and we find
\begin{align}
  \frac{d{\cal I}_y(k_z)}{d \mu}
  =&\, \frac{2 e}{\pi} \int \frac{dk_y}{2 \pi} \mbox{Re}\,
  \frac{\partial \phi_+}{\partial k_y}
  \frac{1}{e^{-i \phi} + 1},
  \label{eq:Iscat4}
\end{align}
where, as before, $\phi(k_y,k_z) = \phi_+(k_y,k_z) - \phi_-(-k_y,-k_z)$. Using $\mbox{Re}\, 1/(e^{-i \phi}+1) = 1/2 - \pi \delta(\phi - \pi)$ we find that $d{\cal I}_y(k_z)/d\mu \sim e \partial \phi_+/\partial k_y$, which is the same order-of-magnitude estimate as one would obtain from Eq.\ (\ref{eq:jy}) by cutting off the small-$d_-$-divergence at $d_- \sim \Delta$. [We note that the condition $d_- \ll \Delta$ may not be fulfilled for all $k_y$ simultaneously, so that, strictly speaking, the approximations leading to Eq.\ (\ref{eq:Iscat4}) do not apply to the full range of the $k_y$-integration. This, however, does not affect the order-of-magnitude estimate of $d{\cal I}_y(k_z)/d\mu \sim e \partial \phi_+/\partial k_y$ that follows from Eq.\ (\ref{eq:Iscat4}).]

We thus find that $d{\cal I}_y(k_z)/d\mu \sim e \partial \phi_+/\partial k_y$ is a regular function of $k_z$ in the vicinity of the Weyl points at $k_z = \pm k_0$. Since the range of momenta $k_z$ affected by the violation of the condition $d_{\tau} \gg \Delta$ is correspondingly small, we conclude that the contribution of the Weyl points to the total current $dI_y/d\mu$ is small and that one may obtain $dI_y/d\mu$ by integration of the $k_z$-resolved result of Eq.\ (\ref{eq:jy}) for $d{\cal I}_y(k_z)/d\mu$, omitting the immediate vicinity of the Weyl points from the integration range.

\begin{figure}
\includegraphics[width=\columnwidth]{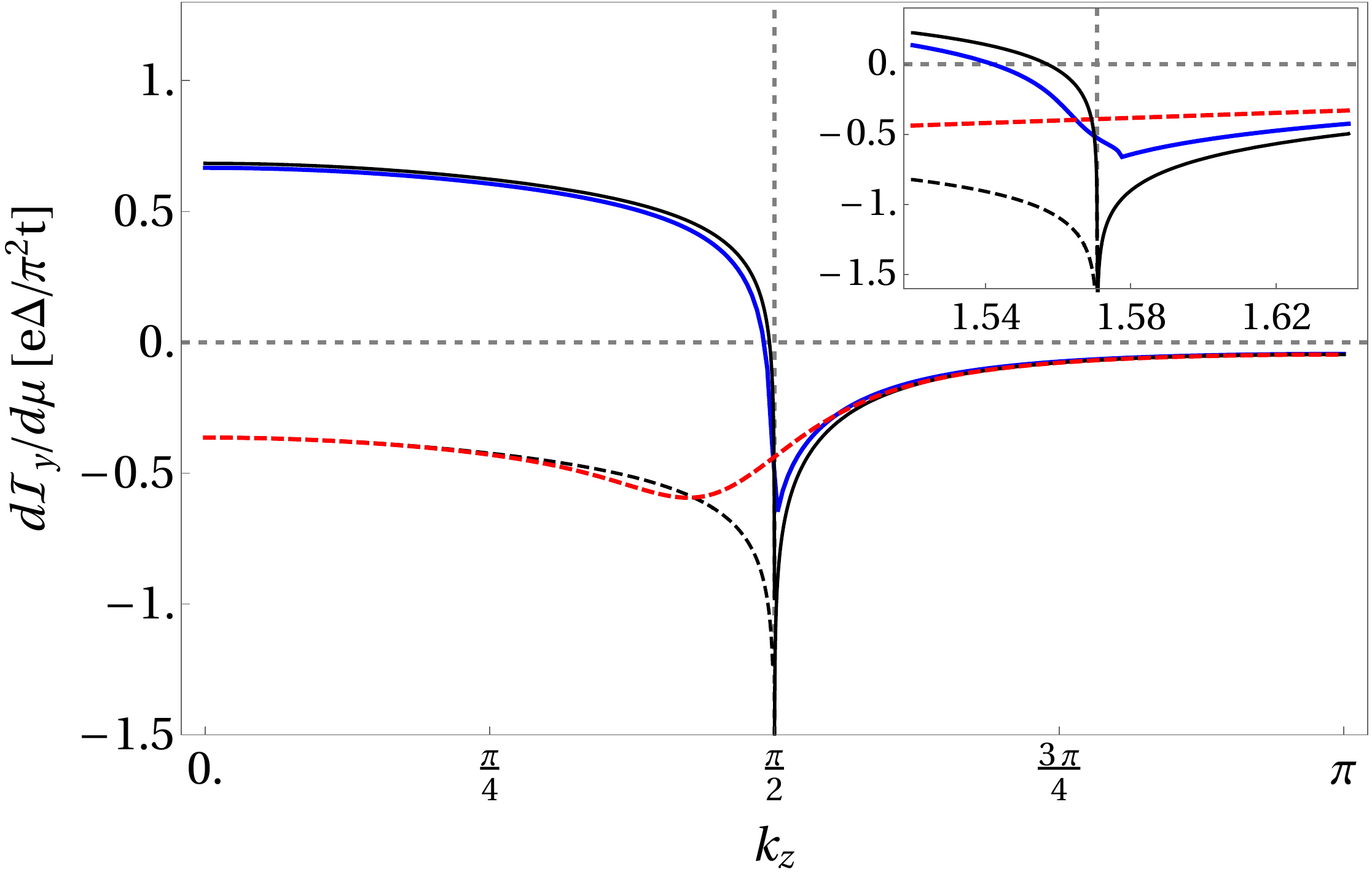}
\caption{$k_z$-resolved equilibrium current $d{\cal I}_y/d\mu$. The superconducting gap $\Delta = 0.01\,  t $; the other parameters are the same as in Fig.\ \ref{fig:analytics}. The solid blue and dashed red curves are obtained from Eq.\ (\ref{eq:Iscat2}) with finite chemical potential $\mu = 0.01\, t$ 
and $\mu = 10^{-6}\, t$, respectively; 
The width of the normal region is 
$W=300$ and $W=5$, respectively. The solid and dashed black curves are obtained from Eqs.\ (\ref{eq:jy}) and (\ref{eq:jyfinite}), respectively. The inset shows a closeup at the Weyl node at $k_0 \approx \pi/2$. The discontinuity in the derivative of $d{\cal I}_y/d\mu$ vs. $k_z$ near $k_0$ is a finite-size effect and disappears upon further increasing $W$.}
\label{fig:kz}
\end{figure}

\subsection{Numerical results}

In Fig.\ \ref{fig:kz} we compare the $k_z$-resolved equilibrium current $d{\cal I}_y(k_z)/d\mu$ obtained directly from Eq.\ \eqref{eq:Iscat2} with the approximation of Eq.\ \eqref{eq:jy}. We find excellent agreement away from the Weyl points. We observe that $d{\cal I}_y(k_z)/d\mu$ has opposite signs for $\mu \ll T_- d_-$ and $\mu \gg T_- d_-$ in the topological regime ($k_z$ between the Weyl points), while there is no difference between the cases of large and small $\mu$ in the trivial regime. Except for the finite-size effect at ultrasmall chemical potentials, we observe only a weak dependence on the width $W$ of the normal region, which is bound to the small vicinity ($d_-\lesssim \Delta$) of Weyl nodes (data not shown). 

\begin{figure}
\includegraphics[width=\columnwidth]{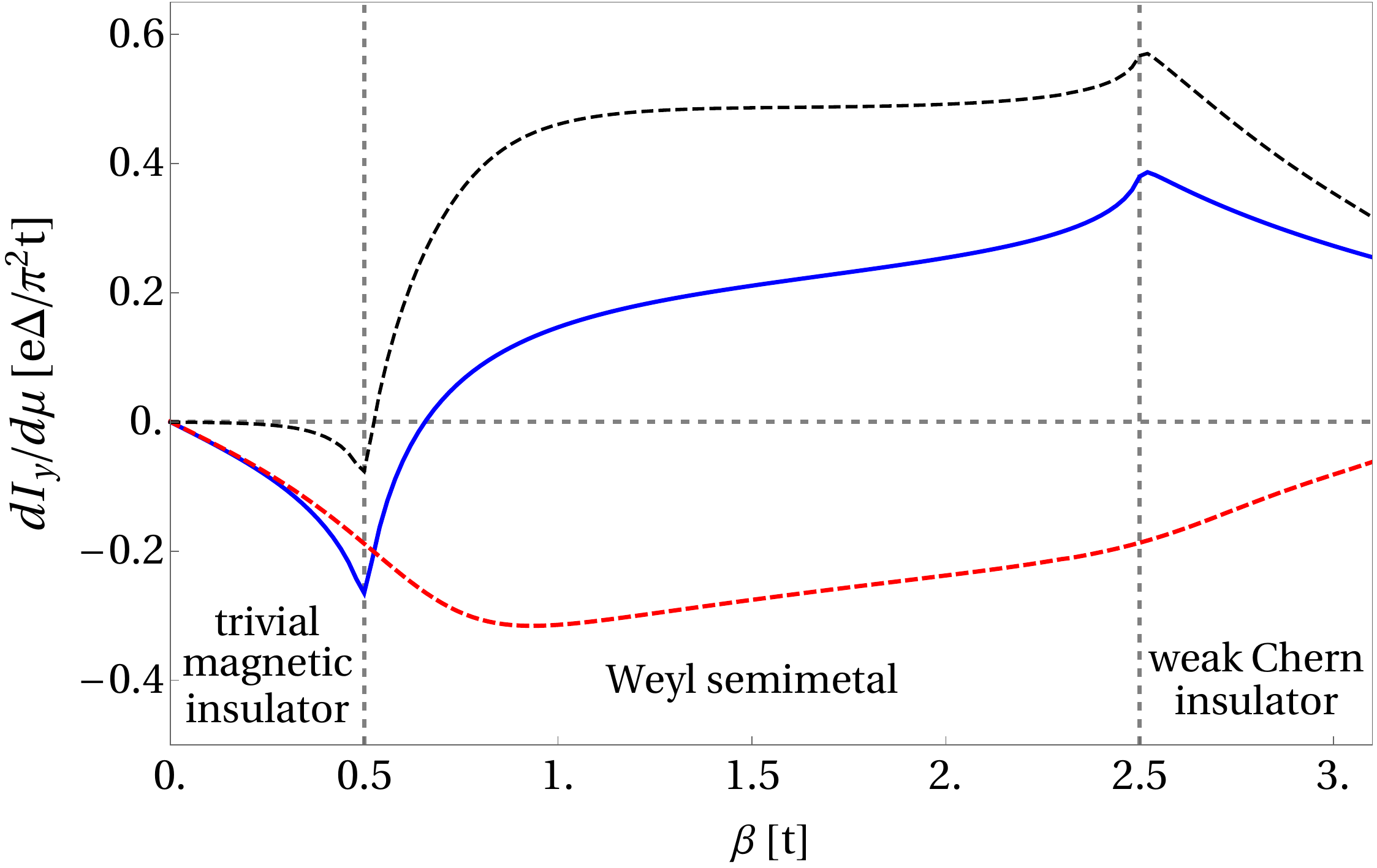}
\caption{Equilibrium current $dI_y/d\mu$ as a function of the exchange field $\beta$.  The solid blue curve is for finite chemical potential $\mu = 0.01\, t$ and 
width $W=300$, which meets the condition $\mu \gg T_- d_-$ for most of reciprocal space. The dashed red curve is for ultrasmall chemical potential 
$\mu = 10^{-6}\, t$ and width $W=5$, which meets the condition $\mu \ll T_- d_-$ for most of reciprocal space. The black dashed curve shows the difference of these two cases, which is the contribution to $dI_y/d\mu$ associated with the Fermi arcs. Other  parameters are same as in Figs.\ \ref{fig:analytics} and \ref{fig:kz}.}
	\label{fig:beta}
\end{figure}

Figure \ref{fig:beta} shows the total current density $dI_y/d\mu$, see Eq.\ (\ref{eq:Itotal}), as a function of the exchange field $\beta$. For comparison, the ultrasmall-$\mu$ limit and the difference between the cases of ultrasmall and finite $\mu$ are also shown (dashed curves in Fig.\ \ref{fig:beta}). The current vanishes at $\beta=0$ because the system is time-reversal invariant there. Its magnitude increases with $\beta$ in the trivial insulator regime $\beta< m_0$. Upon entering the Weyl-semimetal regime, $dI_y/d\mu$ receives an upturn due to the positive contribution of the Fermi arcs. In the weak Chern insulator regime $\beta > m_0 + 2 t_z'$, $dI_y/d\mu$ decreases upon (further) increasing $\beta$, but the difference between ultrasmall and finite chemical potential $\mu$ (dashed curve) persists.

To understand the apparent plateau in the Weyl-semimetal region $m_0 < \beta < m_0 + 2 t_2'$ and the decrease with $\beta$ in the Chern-insulator regime $\beta > m_0 + 2 t_z'$, we note that the order of magnitude of the contribution of Fermi arcs (the difference between $dI_y/d\mu$ for $\mu\gg T_-d_-$ and $\mu \ll T_-d_-$) can be estimated from the difference of Eqs.\ (\ref{tcur}) and (\ref{diff}), multiplying by the distance $2 k_0$ between the Weyl points in the topological region,
\begin{align}
  \frac{  d I^\mathrm{FA}_y}{d\mu} \sim \frac{ e\Delta k_0 }{\beta+m_0},
\label{FA}
\end{align}
where one needs to set $k_0 = \pi$ in the Chern-insulator regime. The apparent plateau in the Weyl-semimetal regime appears, because the increase of the factor $k_0$ in the numerator with $\beta$ is compensated by the increase of the denominator. In the Chern-insulator regime, the numerator in Eq.\ (\ref{FA}) is independent of $\beta$, whereas the denominator continues to increase with $\beta$, explaining the decrease of the current in the Chern-insulator regime. Note that $k_0$ has a singular dependence on $\beta$ at the boundaries of the Weyl-semimetal regime at $\beta=m_0$ and $\beta=m_0+2t'_z$, see Eq.\ \eqref{eq:weylpoints}, which relates to the sharp upturns of the current. We verified that these sharp features are eliminated if $dI_y/d\mu$ is considered as a function of the node separation $2 k_0$ in the Weyl-semimetal regime (data not shown).

\section{Discussion and conclusion
\label{sec:discussion}}

We have investigated the equilibrium current in a minimal model describing an SN heterostructure, where S is a conventional s-wave superconductor and, depending on the value of the exchange field $\beta$, the normal region (N) can be a magnetic insulator with a topologically trivial band structure, a Weyl semimetal with broken time-reversal symmetry, or a three-dimensional weak Chern insulator. The constituents of the heterostructure are  microscopically inversion-symmetric, so that inversion symmetry is broken only by the heterostructure geometry.
In all three regimes, time-reversal symmetry is broken by the exchange field. 

In the trivial-insulator regime 	we find an
 equilibrium current that is proportional to the exchange field $\beta$ at small $\beta$. 
It quantifies the
interface current  of a superconductor - magnetic insulator heterostructure,
which is known to be generally possible in the presence of
spin-orbit coupling. 
Previously such an equilibrium current has been predicted
only for a  system with interfacial Rashba spin-orbit coupling \cite{Mironov2017a}, instead of the intrinsic
spin-orbit coupling considered here.

In the topological regime of a Weyl semimetal or a
weak Chern insulator the current shows a qualitatively 
different behavior. Upon entering the topological regimes the $\beta$-dependence of the equilibrium current abruptly changes, causing a reversal of the sign of the current well inside the topological regime. 
The decisive contribution 
comes from the topological surface states, which we 
can identify within a minimal model 
(motivated by materials of the Bi$_2$Se$_3$ family \cite{Vazifeh2013}) by
comparing the equilibrium currents for  a chemical 
potential inside and above the finite-size gap of the 
surface states. In contrast, the Weyl nodes of the bulk band
structure, which the Fermi arcs connect, do not give a significant contribution to the equilibrium current.
 
That we find a large contribution of Fermi arcs and an insignificant contribution of Weyl nodes relates to previous studies which found that the bulk states of an inversion-symmetric, magnetic Weyl semimetal are mainly unaffected by  superconductivity due to a ``chirality blockade'' \cite{Bovenzi2017}.   Accordingly, we expect that 
this would change if the chirality blockade is 
lifted, which happens when at least one of the constituents 
of the heterostructure  breaks the microscopic inversion symmetry \cite{Bovenzi2017}.
 In our model, the chirality blockade manifests itself through the fact that Andreev reflection from the superconductor switches quasiparticles between the topologically trivial high-energy band and the (potentially)
 topologically nontrivial low-energy band. It is this connection of the trivial and the nontrivial band by the superconducting pairing that also makes the magnitude of the equilibrium current non-universal in both the topologically trivial and nontrivial parameter regimes. 

Whereas the ``chirality blockade'' prevents the bulk Weyl points to be strongly affected by the proximity superconductivity, Fermi-arc surface states at the interface with the superconductor, on the other hand, undergo a renormalization of their effective charge \cite{Baireuther2017}, which however is
weak because of the chirality blockade. Relating the Fermi-arc current contribution of Eq.\ \eqref{FA} to the charge renormalization of Fermi arcs one can interpret the former in terms of an uncompensated chiral current of surface states. Specifically, one can consider that each Fermi arc contributes to the current density
\begin{align}
  \frac{d I_y^{\rm (arc)}}{d\mu} = \mathrm{sign}\,(v) \frac{k_0 q}{(2\pi)^2},
\end{align}
where $v$ is the velocity of the Fermi arc and $q$ the effective charge. The Fermi-arc contribution to the current of the Fermi arcs is reproduced if the charge at the superconductor interface is renormalized to  
\begin{equation}
q \sim -e\big[1-\Delta/(\beta+m_0)\big],
\end{equation}
while the charge of the opposite surface remains unaffected ($q=-e$).  The sign of the Fermi-arc 
velocity has been discussed in 
Sec.\ \ref{sec:model} and is illustrated in Fig.\ \ref{fig:intro}.

The contribution of Fermi arcs can be seen as 
 a real-space counterpart to the superconductivity-enabled equilibrium chiral magnetic effect \cite{Obrien2017, Pacholski2020}, in which a disbalance of
 chiral Landau levels of a pair of Weyl Fermions is 
 produced by current- or flux-biased bulk superconductivity acting asymmetrically in momentum space on the chiral Landau levels. The fundamental connection of  
chiral Landau levels and Fermi arcs allows for the complementary effect that we just described. The 
differences between chiral Landau levels and
 Fermi arcs are that the latter 
 continue to exist in zero magnetic field
and are separated in real space. Our work shows that 
these differences can be used to realize the equilibrium
chiral magnetic effect via the superconducting 
proximity effect, without flux or current bias, and at zero magnetic field. 

Our work, however,  also shows that the experimental 
detection of this effect is challenging because the 
 equilibrium current 
is not exclusively due to Fermi arcs.
The isolation of the 
Fermi-arc contribution that we could obtain in the 
minimal model 
(relying on an ultrasmall chemical potential or an
ultrasmall, constant width of the Weyl semimetal, 
and mirror antisymmetry) 
 does not seem to be experimentally 
realizable on the basis of existing materials. 
We believe, however, that characteristic signatures 
or other peculiar effects
may be found in further studies of the equilibrium current, such as exploring its response to external magnetic fields.

\textit{Acknowledgments}. The authors would like to thank I.\ Adagideli and O.\ Kashuba for valuable discussions.
This research was supported by the German Science Foundation (DFG) through grant no.\ 18688556 and by project A02 of the CRC-TR 183 ``entangled states of matter''.

\appendix

\section{$[d {\cal I}_y(k_z)/d\mu]^{\rm (s)}$ for $\mu \downarrow 0$}
\label{app:a}

To show that the singular contribution to $d {\cal I}_y/d\mu$ changes sign in the limit $\mu \ll T_- d_-$ of an ``ultrasmall'' chemical potential (as compared to the case $\mu \gg T_- d_-$ of a ``finite'' chemical potential), we consider the regime of small $k_y$ and $\mu$ in more detail. The equilibrium current for finite $W$ is found from Eq.\ (\ref{eq:Iscat2}) by replacing $r_+ r_-^*$ by $-e^{i \phi} w^*$, where the function $w(\mu + i \omega - k_y t)$ is given in Eq.\ (\ref{eq:rminfinitesize}), and by restricting the $k_y$-integration to the interval $-\delta < k_y < \delta$,
\begin{align}
  {\cal I}_y(k_z)^{\rm (s)} =&\,
  \frac{2e}{\pi}
  \int_{-\delta}^{\delta} \frac{dk_y}{2 \pi} \mbox{Im}\,
  \int_{0}^{\infty} d\omega
  \frac{\partial \phi_+}{\partial k_y}
  \frac{w^*}{e^{2 i \gamma(\omega) - i \phi} - w^*}.
  \label{eq:Iscat3}
\end{align}
The integration boundaries $\pm \delta$ are chosen such that, on the one hand, $w \approx 1$ for $|k_y| = \delta$, whereas, on the other hand, $\delta \downarrow 0$ as $T_- \to 0$.

To find $[d{\cal I}_y(k_z)/d \mu]^{\rm (s)}$, we have to differentiate the integrand in Eq.\ (\ref{eq:Iscat3}) to $\mu$. Using that for small $k_y$ one has $\partial w/\partial \mu = -(1/t) \partial w/\partial k_y$ and $\partial \phi/\partial \mu = (1/d_+-1/d_-) = -(1/t) \partial \phi/\partial k_y - 2/d_-$ and using that $\phi_+$ is an odd function of $k_y$ for $\mu \to 0$, so that we may treat $\partial \phi_+/\partial k_y$ as a constant inside the integration range $-\delta < k_y < \delta$, we obtain
\begin{align}
  \frac{d{\cal I}_y(k_z)^{\rm (s)}}{d\mu} =&\,
  \frac{2 e}{\pi} \int_{-\delta}^{\delta} \frac{dk_y}{2 \pi}
  \mbox{Im}\,
  \int_{0}^{\infty} d\omega
  \frac{\partial \phi_+}{\partial k_y}
  \\ \nonumber &\, \mbox{} \times
  \left( - \frac{1}{t} \frac{d}{d k_y}
  - \frac{2}{d_-} \frac{\partial}{\partial \phi} \right)
  \frac{w^*}{e^{2 i \gamma(\omega) - i \phi} - w^*}
\end{align}
Since the first term between the brackets, which is proportional to $d/d k_y$, is a total derivative and since $w^* \approx 1$ at both ends of the integration domain, we may set $w^* \to 1$ in the integrand when evaluating the first term. This allows us to relate the first term to the equilibrium current at finite $\mu$. Again using that $(1/t) \partial \phi/\partial k_y = -(1/d_++1/d_-) = (d_++d_-)/(d_+-d_-) \partial \phi/\partial \mu$, we recognize that the first term is $-(d_++d_-)/(d_+-d_-)$ times the singular contribution of Eq.\ (\ref{eq:jys}). 

The second term between the brackets
vanishes to leading order in $\Delta/d_-$:
To leading order in $\Delta/d_-$ the energy dependence
in $w^*$ can be neglected and the $\omega$ 
integration can be performed similarly as when going from Eq.\ 
\eqref{eq:Iscat2} to Eq.\ \eqref{eq:Iphi} with the phase modified by $w^*$,
which approaches $1$ upon taking
 the limit $T_-\to 0$. The whole integrand is  
thus non-singular in this limit and, upon integration, the term vanishes for $T_-\to 0$ due to  the vanishing integration range.

\section{Continuity of the current in the limit $T_- \downarrow 0$}
\label{app:b}

In the main text we derived the current 
at the transmission
amplitude set to zero from the beginning. 
Here we repeat the calculation in a more careful way, 
taking the limit $T_-\to 0$ at the end, to 
show that the 
current is a continuous function of $T_-$ at $T_-=0$.
For simplicity we only consider the well-established topological regimes 
at $k_z=0$, $\beta=m_0+t$, and $t=t'=t_z'$. The goal is 
thus to reproduce Eqs.\ \eqref{tcur} and \eqref{diff}.

Starting point is Eq.\ \eqref{eq:Iscat2}, where we set $k_z=0$,
\begin{align}
  \label{eq:Iscata}
  {\cal I}_y(0) =&\, \frac{2 e}{\pi} \int \frac{dk_y}{2 \pi}
  \mbox{Re}\, \int_0^{\infty} d\omega 
  \frac{\partial r_+(i \omega;k_y,0)}{\partial k_y}
  \\ \nonumber &\, \mbox{} \times
  \frac{r_-(i \omega;-k_y,0)^*}{e^{2 i \gamma(i \omega)} + r_+(i \omega;k_y,0) r_-(i \omega;-k_y,0)^*}.
\end{align}
We consider leading order in the 
gap $d_+\approx \beta+m_0$ of the high-energy band,
allowing to 
approximate $r_+^\mathrm{(W)} = i \exp[-i t\sin k_y /(\beta+m_0)]$ and leading to
\begin{align}
  {\cal I}_y(0) =&\, \frac{2 e t}{\pi(\beta+m_0)} \int \frac{dk_y}{2 \pi}\cos k_y\; 
  \mbox{Re}\, \int_0^{\infty} d\omega 
  \\ \nonumber &\, \mbox{} \times
  \frac{r_-(i \omega;-k_y,0)^*}{e^{2 i \gamma(i \omega)} + i r_-(i \omega;-k_y,0)^*}.
\end{align}
For the non-trivial band we take the full reflection 
amplitude of Eq.\ \eqref{eq:rtau},
\begin{equation}
  r_- = -i e^{i \phi_-}
  \frac{e^{i \phi_-'} + \sqrt{1 - T}}
       {e^{i \phi_-'} \sqrt{1 - T} + 1},
       \label{eq:rtaua}
\end{equation}
where for brevity we have written $T$ instead of $T_-$.
In the well-established topological regime at $k_z=0$, $\beta=m_0+t$, and $t=t'=t_z'$, the reflection phase for the non-trivial band is $\phi_-(k_y,k_z)=\pi+\mu/t - k_y$.
Further, we introduce
 $Z=\exp(-i k_y)$ and  
 use $dk_y\cos k_y = i dZ(1+Z^2)/2Z^2$, as well as $\omega = \Delta \sinh{\zeta}$ and $d\omega = d\zeta \Delta \cosh{\zeta}$ (so that $e^{2i \gamma} = -e^{2 \zeta}$) to obtain
\begin{align}
{\cal I}_y(0)
=&  -\frac{\Delta \, e \, t}{\pi (\beta+m_0) \sqrt{1-T}}\mathrm{Re}\; 
 \int_0^\infty d\zeta  \oint\frac{dZ}{2\pi i}\cosh{\zeta} \nonumber\\
 &\;\;\; \times
\frac{i(e^{-i \mu/t}-\sqrt{1-T}Z)(1+Z^2)}{Z(Z-Z_-)(Z-Z_+)}, \label{eq:apjyintegral}
\end{align}
where
\begin{align}
Z_\pm = e^{\zeta}\frac{\pm i \sqrt{\sin^2(i\zeta-\mu/t)-T}
+ \cos( i\zeta-\mu/t) }{\sqrt{1-T}} .
\end{align}
The integration contour of 
 $Z$ is the unit circle in the complex plane
 enclosing two poles, one at $Z=0$ and the other at $Z=Z_+$. 

For $T=0$ only the pole at $Z=0$ contributes to the integral, due to
cancellation of the $(Z-Z_+)$ term of the denominator with the first term
of the numerator in Eq.\ \eqref{eq:apjyintegral}, and it gives
\begin{align}
{\cal I}_y^{(0)}(0) =&- \frac{\Delta\, e\, t}{\pi (\beta+m_0)}\mathrm{Im}\; 
 \int_0^\infty d\zeta \cosh{\zeta}   \;
e^{-2\zeta -i\mu/t}, \label{eq:apjy0}
\end{align}
which for $ \mu \ll t$ evaluates to
\begin{equation}
\frac{d {\cal I}_y^{(0)}(0) }{d\mu} = \frac{2e\Delta}{3\pi (\beta+m_0) }, 
\end{equation}
reproducing Eq.\ \eqref{tcur}.

For $T>0$ both poles at $Z=0$ and $Z=Z_+$ contribute to the integration. The contribution of the $Z=0$ pole gives the same as the result Eq.\ \eqref{eq:apjy0} for $T=0$ up to a factor of $1/\sqrt{1-T} \to 1$. 

The contribution to the integral from the pole at $Z=Z_+$ is
\begin{align}
{\cal I}_y^{(1)}(0) 	
=& 	 -\frac{ e\Delta}{2 \pi (\beta+m_0)}\mathrm{Im}\; 
	\int_0^\infty d\zeta
	\, g(i\zeta-\mu/t) \nonumber \\ &\, \times \left[z(i\zeta-\mu/t)
 \left(1+e^{2\zeta}\right) \right. \nonumber \\ &\, + \left.  z^{-1}(i\zeta-\mu/t)\left(1+e^{-2\zeta}\right)\right]
	\label{eq:apjy1},
\end{align}
where we abbreviated
\begin{align}
& g(i\zeta-\mu/t)=  \frac{e^{-i\mu/t}-\sqrt{1-T}Z_+}{\sqrt{1-T}(Z_+ - Z_-)}, \\
& z(i\zeta-\mu/t)= e^{-\zeta}Z_+.
\end{align}
(One verifies that $g$ and $z$ are functions of $i \zeta - \mu/t$ only.)
Since it contributes for $T > 0$ only, the pole at $Z_+$ can be seen to represent a contribution to the equilibrium current from the Fermi arc at the insulating side of the semimetal. To estimate this contribution in the limit of small $T$, we note that the difference $Z_+ - Z_-$ is
\begin{align}
Z_+ - Z_- = 2ie^{\zeta} \sqrt{\frac{\sin^2{(i\zeta-\mu/t)}-T}{1-T}} .
\end{align}
To further evaluate this expression in the limit of small transmission $T$, we note that for $T \ll 1$ one has
\begin{align}
	Z_+ = e^{-i\mu/t} \left[ 1-i\frac{T}{2} \cot{(i\zeta-\mu/t)} + \ldots \right] . \label{eq:apZmexp}
\end{align}
In the limit of large $\zeta$, this expansion is convergent and gives a numerator of order $T$ in Eq.\ \eqref{eq:apjy1}. Hence, for large $\zeta$, the integral in Eq.\ \eqref{eq:apjy1} is convergent and of order $T$. If $\mu \neq 0$ this conclusion applies to the entire integration domain $\zeta > 0$, so that we conclude that the finite-$T$ correction to the result shown in Eq.\ \eqref{eq:apjy0} is of order $T$ and smoothly vanishes for $T \downarrow 0$ if $\mu \neq 0$. The case $\mu = 0$ is different because then the expansion shown in Eq.\ \eqref{eq:apZmexp} is singular for $\zeta \to 0$. In the limit of small $\zeta$ one finds, if $\mu = 0$, that
\begin{align}
 g(i\zeta)
  =&\,
-  \frac{\sqrt{\zeta^2 + T} - \zeta}{2 \sqrt{\zeta^2 + T}} \nonumber \\ =&\,-
  \frac{T}{2\sqrt{\zeta^2 + T}
    (\sqrt{\zeta^2 + T} + \zeta )}.
  \label{eq:fraction}
\end{align}
We now divide up the $\zeta$ integral into a region $0 < \zeta < T^{\alpha/4}$ and a region $T^{\alpha} < \zeta$ with $0 < \alpha < 1/2$. In the former region, the remaining factors of the integration are approximately constant and integration of Eq.\ \eqref{eq:fraction} gives a contribution to ${\cal I}_y^{(1)}(0)$ that is of order $\sqrt{T}$. In the region $\zeta > T^{\alpha}$ one may still use the small-$T$ expansion from Eq.\ \eqref{eq:apZmexp} to arrive at a systematic expansion around the result at $T=0$. Since both contributions to the integral vanish in the limit $T \to 0$, we conclude that ${\cal I}_y^{(1)}(0) 	\to 0$ for $T \to 0$ even if $\mu = 0$, although the convergence may be slower than for generic $\mu$.

We now consider the derivative of \eqref{eq:apjy1}
 with respect to $\mu$ at
$\mu=0$ before taking the limit $T\to 0$. 
We use that
 $d/ d\mu =( i/t) d/d\zeta$ acting on $ z(i\zeta-\mu/t)$
 and $ g(i\zeta-\mu/t)$, to obtain
 \begin{align}
\frac{d {\cal I}_y^{(1)}(0) }{d\mu}	
=& 	 -\frac{e \Delta }{2 \pi (\beta+m_0) d_-}\mathrm{Re}\; 
	\int_0^\infty d\zeta \Big(1+e^{2\zeta}\Big)\nonumber\\
& \times
	\frac{d }{d\zeta} g(i\zeta)z(i\zeta) 	+ \Big(1+e^{-2\zeta}\Big) \frac{d }{d\zeta}\frac{g(i\zeta)}{z(i\zeta)}.
\end{align}
Using 
\begin{align}
\lim_{T\to 0} \frac{g(0)}{z(0)} = \lim_{T\to 0} g(0)z(0) 
=-\frac{1}{2} ,
\end{align}
partial integration gives,
 \begin{align}
\frac{d {\cal I}_y^{(1)}(0) }{d\mu}	
=& 	 -\frac{ e \Delta}{\pi (\beta+m_0) }\bigg[1-
\mathrm{Re}\; 
	\int_0^\infty d\zeta \Big( e^{2\zeta} g(i\zeta)z(i\zeta)
	\nonumber \\
&	-e^{-2\zeta} \frac{g(i\zeta)}{z(i\zeta)}\Big)\bigg].
\end{align}
The remaining integral vanishes for $T\to 0$ similarly  as 
the current in   \eqref{eq:apjy1}  at $\mu=0$ as 
shown above, hence 
 \begin{align}
\frac{d{\cal I}_y^{(1)}(0) }{d\mu}	
=& 	 -\frac{ e \Delta}{\pi (\beta+m_0)}.
\end{align}
Thus for the total current
${\cal I}_y^{(0)}(0) +{\cal I}_y^{(1)}(0) $ in the ordered limit $\mu\to 0$, $T\to 0$ we obtain
 \begin{align}
\frac{d{\cal I}_y(0)  }{d\mu}	
=& 	 -\frac{e  \Delta}{3 \pi (\beta+m_0) },
\end{align}
reproducing Eq.\ \eqref{diff}.

\bibliography{library}

\end{document}